\documentclass{article}

\usepackage{PRIMEarxiv}

\usepackage[utf8]{inputenc} % allow utf-8 input
\usepackage[T1]{fontenc}    % use 8-bit T1 fonts
\usepackage{hyperref}       % hyperlinks
\usepackage{url}            % simple URL typesetting
\usepackage{booktabs}       % professional-quality tables
\usepackage{amsfonts}       % blackboard math symbols
\usepackage{nicefrac}       % compact symbols for 1/2, etc.
\usepackage{microtype}      % microtypography
\usepackage{lipsum}
\usepackage{subcaption}
\usepackage{fancyhdr}       % header
\usepackage{graphicx}       % graphics
\graphicspath{{media/}}     % organize your images and other figures under media/ folder
\usepackage{amsmath}

\expandafter\def\expandafter\normalsize\expandafter{%
    \normalsize
    \setlength\abovedisplayskip{5pt}
    \setlength\belowdisplayskip{5pt}
    \setlength\abovedisplayshortskip{5pt}
    \setlength\belowdisplayshortskip{5pt}
}
\usepackage{caption} 
\title{Multi-task learning for tissue segmentation and tumor detection in colorectal cancer histology slides}
\author{
  Lydia A. Schoenpflug, Maxime W. Lafarge, Anja L. Frei, Viktor H. Koelzer\\
  Department of Pathology and Molecular Pathology\\
  University Hospital and University of Zurich \\
  Zurich, Switzerland\\
}

\begin{document}
\maketitle
\begin{abstract}
Automating tissue segmentation and tumor detection in histopathology images of colorectal cancer (CRC) is an enabler for faster diagnostic pathology workflows. At the same time it is a challenging task due to low availability of public annotated datasets and high variability of image appearance. The semi-supervised learning for CRC detection (SemiCOL) challenge 2023 provides partially annotated data to encourage the development of automated solutions for tissue segmentation and tumor detection. We propose a U-Net based multi-task model combined with channel-wise and image-statistics-based color augmentations, as well as test-time augmentation, as a candidate solution to the SemiCOL challenge. Our approach achieved a multi-task Dice score of .8655 (Arm 1) and .8515 (Arm 2) for tissue segmentation and AUROC of .9725  (Arm 1) and 0.9750 (Arm 2) for tumor detection on the challenge validation set.  The source code for our approach is made publicly available at \url{https://github.com/lely475/CTPLab_SemiCOL2023}.
\end{abstract}

% keywords can be removed
%\keywords{Multi-task learning \and Second keyword \and More}

\section{Introduction}
\begin{figure}[htb]
  \centering
  \includegraphics[width=\textwidth]{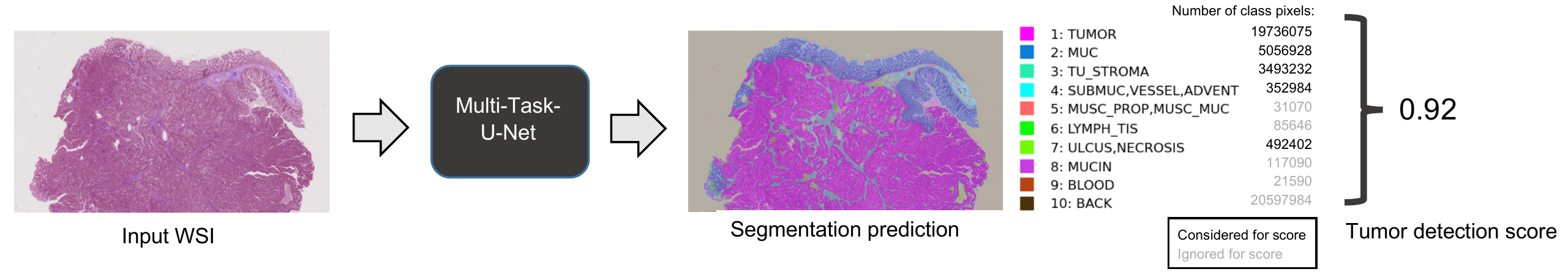}
  \caption{Proposed approach to tissue segmentation and tumor detection: we train a multi-task U-Net-based model for segmentation and classification, but use only the segmentation branch during inference. The tumor detection score is computed based on the number of predicted class pixels, as defined in Equation \ref{eq:tum-det-score}.}
  \label{fig:key-pipeline}
\end{figure}

Colorectal cancer is a leading cause of cancer-related deaths \cite{Bray2018}. One aim of digital pathology is the automation of routine tasks, such as the analysis of tissue for the presence of epithelial tumor tissue in biopsies and bowel resections. The SemiCOL challenge \cite{semicol-challenge} encourages the development of a digital pathology pipeline for tissue segmentation and tumor detection in hematoxylin and eosin (H\&E) stained slides of CRC tissue. A smaller set of data with partial segmentation annotation and a larger set of weakly, slide-level labeled data (tumor present: yes/no) are provided \cite{semicol-data}. Our approach, a tile-based multi-task model, leverages both the segmentation and weakly annotated data during training, with data augmentation for better generalization. Our desired outcome is a model capable of robust tissue segmentation that can be used to derive slide-level tumor detection labels, as shown in Figure \ref{fig:key-pipeline}. We developed a baseline algorithm, where training hyperparameters were chosen based on the performance on an internal validation subset of the challenge training set and locked for all subsequent experiments. We made a comparative analysis of variations building on this baseline model and selected the best performing configuration with regards to an internal validation set.

\section{Model Architecture}
\label{sec:model architecture}
\begin{figure}[htb]
  \centering
  \includegraphics[width=0.85\textwidth]{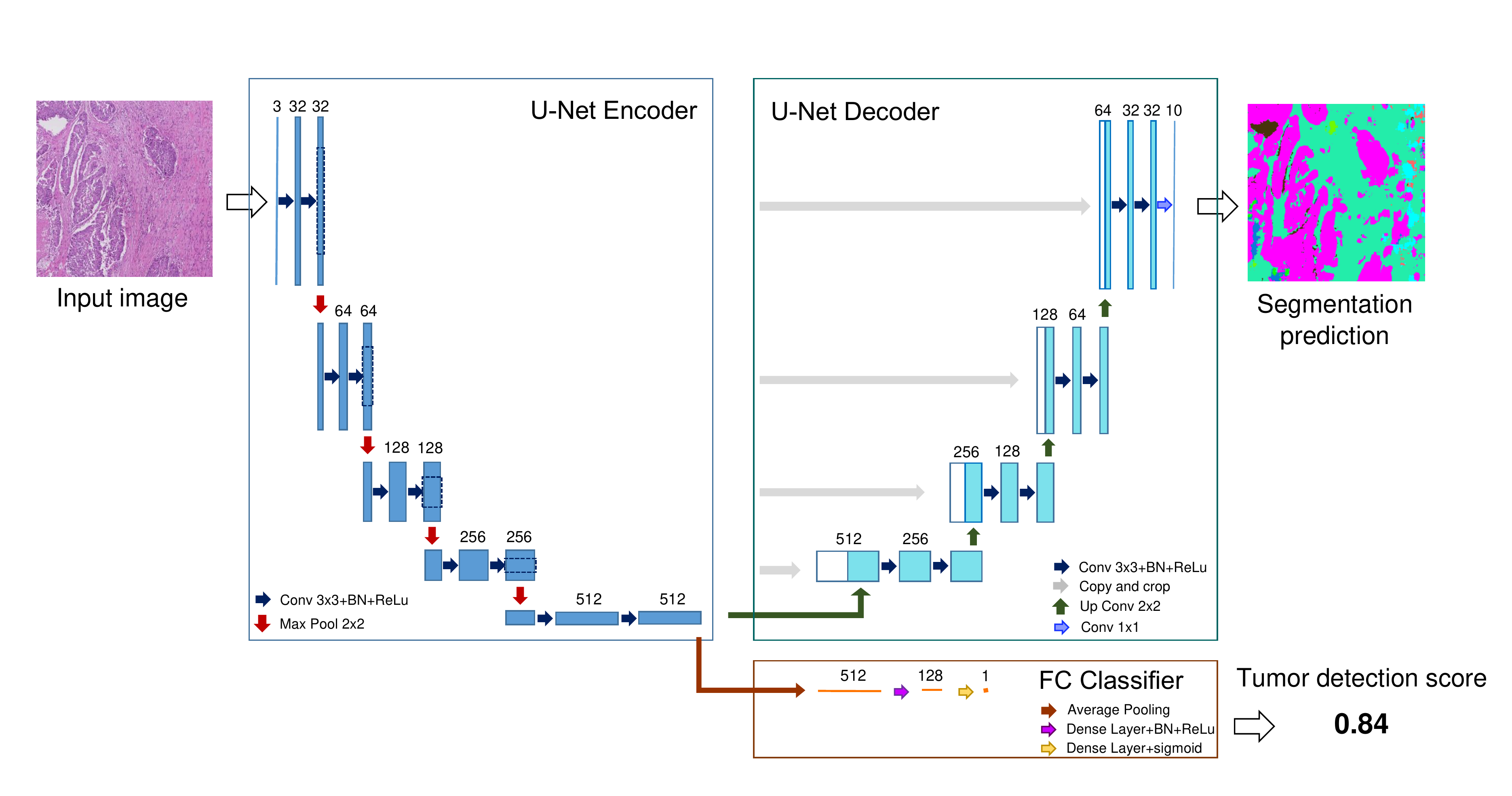}
  \caption{Multi-task model architecture: U-Net based encoder and decoder head for tissue segmentation and fully connected (FC) classifier head for weak supervision in the tumor detection task.}
  \label{fig:arch}
\end{figure}

The chosen model architecture, visualized in Figure \ref{fig:arch}, is based on a U-Net \cite{UNet} architecture with an encoder for feature extraction, followed by a decoder segmentation head for tissue segmentation and a fully connected classifier head for tumor detection. Convolutional layers were implemented without padding, resulting in an output prediction map which is smaller than the input image. For validation and inference, the input is padded by reflecting the border areas, enabling a full comparison of the prediction with its ground-truth. The classifier head is used only during training to enable the use of weakly annotated samples in a weakly supervised fashion. For inference, a tumor detection score is computed based on the segmentation prediction. The model consists of 32 feature channels for the input, resulting in 512 feature channels for the final encoding layer.

\section{Dataset Preparation and Partitioning}
The SemiCOL challenge provides a training and validation set \cite{semicol-data}. 
The SemiCOL training set was split into an internal training and internal validation set on a 80:20 slide-level split with stratified domain and annotation distribution (segmentation training set: 59611 patches from 16 whole slides images (WSIs), segmentation validation set: 12834 patches from 4 WSIs, weakly annotated training set: 399 WSIs, weakly annotated validation set: 100 WSIs). The segmentation set is comprised of patches with a size of 3000x3000 pixels, scanned at magnification 10x, and their respective ground-truth segmentation masks. The weakly annotated set contains 499 slides scanned at magnification 20x.
We homogenized the dataset by downscaling both segmentation patches and weakly annotated slides to a magnification of 5x and extracting tiles of size 300x300 pixels. Due to the small amount of segmentation annotations, the segmentation sets were tiled with 50\% overlap and tiles were kept if at least 1\% of pixels in the tile were annotated. The weakly annotated slides were tiled without overlap and tiles were kept, if they contained at least 50\% tissue, to avoid over-representation of the slide background. The slide-level tumor detection label was assigned to each slide. 
The SemiCOL validation data set, further referred to as the external validation set, consists of one set for evaluating tissue segmentation (858 patches originating from 9 WSIs) and a second set for evaluating tumor detection performance (40 WSIs).

\section{Training Procedure}
\begin{figure}[htb]
\centering
\begin{subfigure}[t]{0.4\textwidth}
  \centering
  \includegraphics[width=0.87\textwidth]{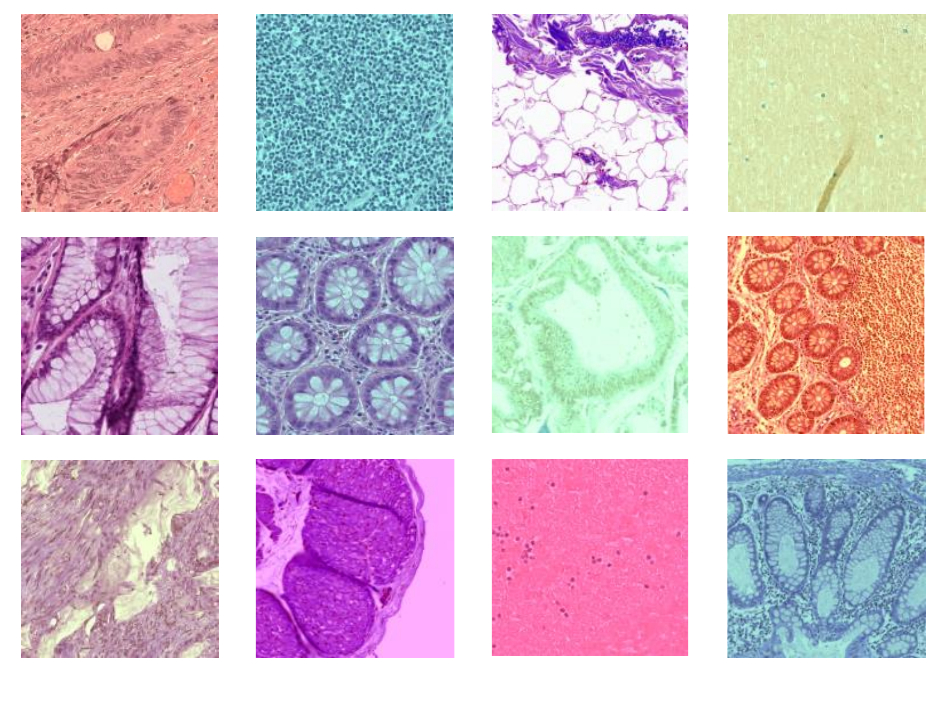}
  \caption{Channel-wise brightness and contrast variation}
  \label{fig:sub1}
\end{subfigure}%
\begin{subfigure}[t]{0.55\textwidth}
  \centering
  \includegraphics[width=0.87\textwidth]{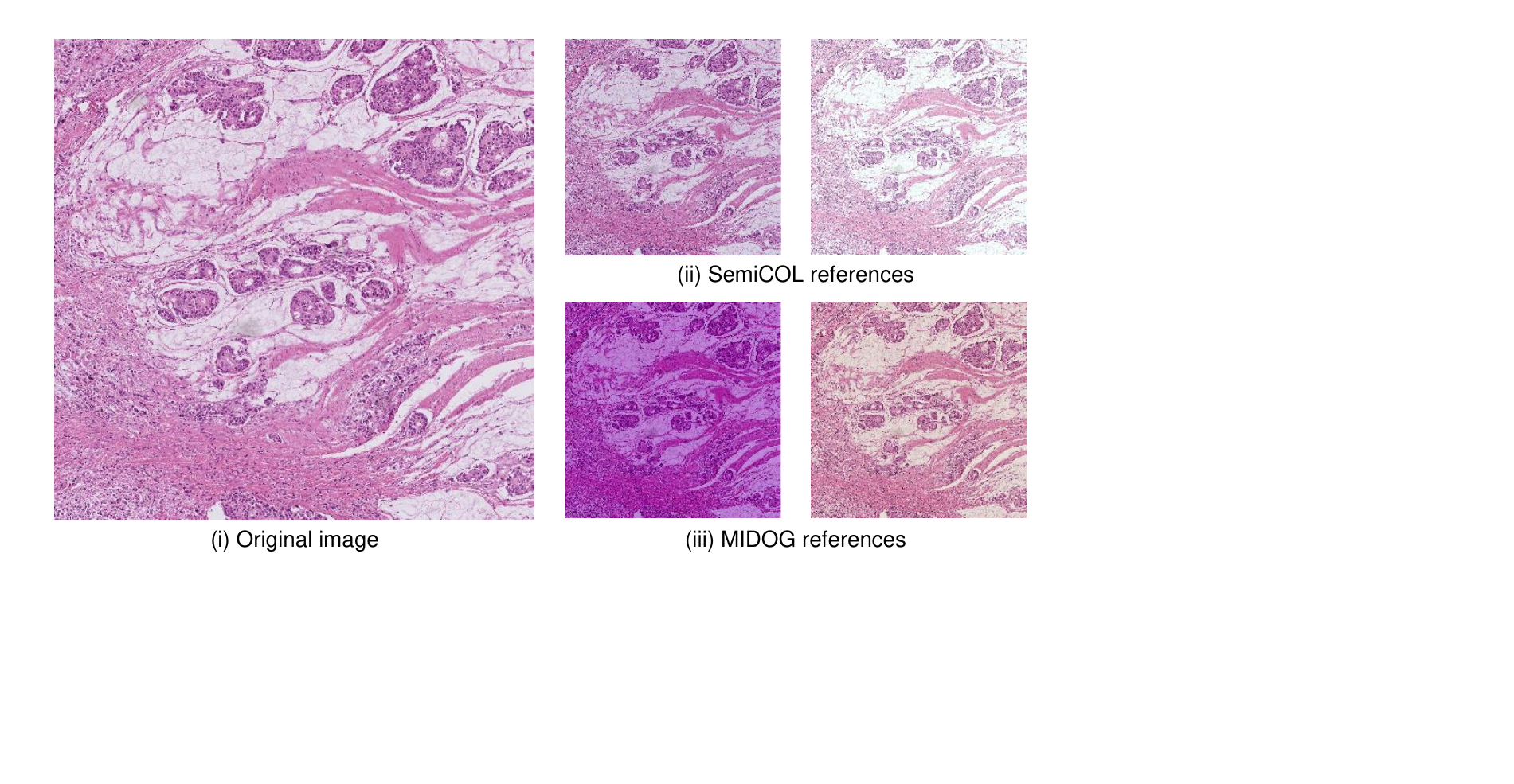}
  \caption{Image-statistics based color augmentation}
  \label{fig:sub2}
\end{subfigure}
\caption{Examples for (a) Channel-wise brightness and contrast variation and (b) Image-statistics based color augmentation, (i): Original image, (ii), (iii): Examples for mean and standard deviation transfer for references from (ii) SemiCOL and (iii) MIDOG dataset.}
\label{fig:augmentation}
\end{figure}

After running an initial set of experiments with a baseline model, we selected the following hyperparameters:
each model is trained for 100 epochs with batch size 128 and stochastic gradient descent with nesterov momentum and weight decay (inital learning rate: 0.2, momentum: 0.9, weight decay: $5\cdot 10^{-6}$) and an exponential learning rate decay with $\gamma=0.97$. For performance evaluation on the internal validation set the model was trained on the internal training set, for evaluation on the external validation set we trained on the full challenge training set, comprised of the combined internal training and validation set.
Loss consists of a cross-entropy loss for segmentation $\text{CE}_{\text{sgm}}$ and a binary cross-entropy loss for tumor detection $\text{BCE}_{\text{td}}$ that are combined in a weighted multi-task loss with $w=0.5$:
\begin{equation}
    \text{Loss} = w \cdot \text{CE}_{\text{sgm}} + (1-w)\cdot \text{BCE}_{\text{td}}
\end{equation}
Each training batch is balanced between samples from the segmentation and the weakly annotated training sets. Tiles without segmentation annotation are ignored for segmentation loss. Tiles with segmentation annotation are assigned the tumor label if the tumor tissue class is present in the segmentation mask, otherwise they are labeled as non-tumorous. Here, a training epoch is defined as the number of batches necessary for the model to see all the tiles of the segmentation training set. The tiles from the weakly annotated training set are randomly undersampled to match the amount of segmentation training tiles.
To improve model robustness we introduced flipping, transposing and random rotation by 90°, 180° or 270° as random geometric augmentations, as well as scale variation by $\pm 10\%$ and random crop to the desired input size of 260x260 pixels. For the baseline model, we applied a random brightness and contrast variation by $\pm 30\%$. All augmentations were applied with a probability of 70\%. 
To achieve better domain generalization, we investigate two further augmentation methods:
\begin{enumerate}
    \item Channel-wise brightness and contrast variation by $\pm 20\%$: this represents many possible colors and shades, thus forcing the model to learn relevant morphological features that are invariant to uninformative color variations \cite{Lafarge-Domain-Invariant, TELLEZ2019101544}. Figure \ref{fig:sub1}  shows the effect on tiles from the SemiCOL training set.  
    \item Image-statistics-based color augmentation: This augmentation protocol is inspired by one of the winning methods \cite{10.1007/978-3-030-97281-3_14} of the mitosis domain generalization in histopathology images (MIDOG) challenge 2022 \cite{AUBREVILLE2023102699, MiDOG}. The authors of \cite{10.1007/978-3-030-97281-3_14} proposed swapping the low frequency component between a given input image and a reference image. The effect can be interpreted as a straight-forward form of inter-scanner style transfer. Since exchanging the lowest frequency is closely equivalent to exchanging the mean pixel value of the image, we investigate a mean exchange as an even more computationally-efficient method.
    The reference means were computed from 2000 randomly sampled images for each scanner type and institution origin, resulting in 10 different mean values from the SemiCOL dataset.
    Additionally, the reference standard deviation was computed and used to re-scale the images to simulate the domain-specific contrast spectrum.
    For further enriching the references, we computed the means and standard deviation of representative domains from the MIDOG challenge 2022. Examples of augmented tiles are shown in Figure \ref{fig:sub2}.
\end{enumerate}

\section{Inference Pipeline} 
For inference, the input WSI is rescaled to magnification 5x and padded by reflecting the boundary regions, so that the unpadded input and predicted output are of identical size.
The rescaled WSI is tiled, and each tile is fed to the model for tissue segmentation. 
The amount of predicted pixels for each segmentation class is then utilized to compute a tumor detection score.
We evaluated different score definitions on the internal validation set. Equation \ref{eq:tum-det-score} shows the selected, best performing score: 
\begin{equation}
    \text{Tumor detection score} = \frac{\text{tumor}+\text{tumor stroma}+\text{ulcus necrosis}}{\text{tumor}+\text{tumor stroma}+\text{ulcus necrosis}+\text{benign mucosa}+ \text{submucosa}}
    \label{eq:tum-det-score}
\end{equation}
where each value refers to the amount of predicted pixels for the class.

Additionally, we evaluated the effect of geometrical test time augmentation for more robust predictions, consisting of all 8 possible rotation and flip combinations. 

\section{Discussion and Results}
Table \ref{tab:table} details the performance on the internal and external validation set for seven experimental settings. 
Arm 1 of the SemiCOL challenge refers to methods that only utilize the datasets provided by the SemiCOL challenge, in Arm 2 additional data can be used.
\begin{table}[htb]
 \caption{Comparison of multi-class Dice score for tissue segmentation and AUROC for tumor detection on the internal and external validation set for our trained models. We start with the baseline model with only the segmentation branch and successively compare to the investigated methods. In Arm 1, only references from the SemiCOL training set are considered for image-statistics-based color augmentation, for Arm 2 MIDOG references are added. Performance on the internal validation set is reported as the average and standard deviation of three runs.}
  \centering
  \begin{tabular}{lcccc}
    \midrule
    \cmidrule(r){1-2}
    & \multicolumn{2}{c}{Internal validation set} & \multicolumn{2}{c}{External validation set} \\
     & Multi-class Dice Score & AUROC & Multi-class Dice Score & AUROC \\
     \midrule
    Only segmentation        & .9152$\pm$.0014 & .8716$\pm$.0106& .6071 & .6350 \\
    + tumor detection branch & .9063$\pm$.0047& .8809$\pm$.0291& .7068 & .7325 \\
    + channel-wise color augmentation & .9287$\pm$.0026 & \textbf{.9379$\pm$.0134} & .8011 & .9425 \\
    \textbf{Arm 1} & & & & \\
    + image-statistics color augmentation & .9329$\pm$.0024 & .9371$\pm$.0103 & .8550 & .9700 \\
    + test-time-augmentation & .9400$\pm$.0020 & .9339$\pm$.0103 & \textbf{.8655} & .9725\\
    \textbf{Arm 2} & & & & \\
    + image-statistics color augmentation & .9351$\pm$.0003 & .9059$\pm$.0245 & .8435 & \textbf{.9750} \\
    + test-time-augmentation & \textbf{.9411$\pm$.0013} & .9025$\pm$.0249 & .8515 & \textbf{.9750} \\
    \bottomrule
  \end{tabular}
  \label{tab:table}
\end{table}

We selected the best performing model based on the internal validation set multi-class Dice score. For both Arm 1 and Arm 2 the top performing configuration is a multi-task model utilizing the tumor detection branch, channel-wise and image-statistics-based color augmentation and test-time-augmentation. It achieved .9400 (Arm 1) and 0.9411 (Arm 2) multi-class Dice score and .9339 (Arm 1) and 0.9025 (Arm 2) AUROC.
The results on the external validation set align with this choice, with .8655 (Arm 1) and 0.8515 (Arm 2) multi-class Dice score and .9725 (Arm 1) and 0.9750 (Arm 2) AUROC.
%We report overall high scores on the internal validation set with minimum .9063 for Dice score and .8716 for AUROC. %The biggest improvement is achieved by adding channel-wise color augmentation.%While the multi-class Dice score can be further improved by adding image-statistics color augmentation and test-time augmentation, the AUROC stays in a similar range for the added methods in Arm 1, and decreases by approximately 3\% for Arm 2.
%The high performance on the internal validation set, in comparison to the lower performance for the same configuration on the external validation set, suggests overfitting. %As slides from the internal validation and training sets can originate from the same scanner types and institutes, the internal validation sets are less expressive with regards to model generalization. 
%The small size of the internal segmentation validation set, consisting of only 4 slides, further aggravates this issue. With regards to the weakly annotated data, the internal validation data set is solely comprised of slides showing tissue from bowel resections, identical to the internal training set. In contrast, the external validation set shows biopsy slides, thus providing a more challenging and expressive set for evaluating model generalization.
%For this reason, we mainly rely on the external validation set for model comparison and selection. %This is possible, due to the very small internal segmentation validation set, consisting of only 4 slides that originate from the same scanner types and institutes as the internal training set. F

Notably, we see an improvement of the multi-class Dice score and AUROC on the external validation set for each method that is added to the baseline. Utilizing the weakly annotated set by considering the tumor detection branch showed an improvement of .0997 for multi-class Dice score and .0975 for AUROC. Therefore employing a multi-class model approach lead to a moderate generalization improvement.
We observe the strongest model improvement by adding channel-wise color augmentation, which increased the multi-class Dice score by .0943 and AUROC by .21. This reflects the effectiveness of color augmentation to enable generalization to different domains such as scanner type and staining protocols, as previously reported in the literature \cite{Lafarge-Domain-Invariant, TELLEZ2019101544}.
While the image-statistics-based color augmentation had a similar effect, it is less significant with an increase of .0539 (Arm 1) and .0424 (Arm 2) for multi-class Dice and .0275 (Arm 1) and 0.0325 (Arm 2) for AUROC. We expected an improved model generalization when adding the MIDOG references for image-statistics-based color augmentation in Arm 2, compared to only using the SemiCOL references in Arm 1, however we observe a higher multi-class Dice score for Arm 1. It remains to be seen, whether the added references in Arm 2 will lead to an improved generalization compared to Arm 1 on the SemiCOL challenge test set.
Lastly, while test-time augmentation leads to an additional, small improvement of approximately .01 for multi-class Dice score for Arm 1 and Arm 2, AUROC stays approximately the same.

%The best performing configuration for Arm 1 and Arm 2 is a multi-task model with channel-wise color augmentation, image-statistics-based color augmentation and test-time augmentation. It achieved .8655 (Arm 1) and 0.8515 (Arm 2) multi-class Dice score and .9725 (Arm 1) and 0.9750 (Arm 2) AUROC on the external validation set.
For future work, we suggest an extension of the training set with active learning, where additional annotations are provided by an expert pathologist.
Further improvement might be achieved by curating the tiles selected from the weakly annotated set, instead of randomly sampling them. %The curation can be achieved by predicting on the weakly annotated slides with our current model and selecting tiles with high model prediction confidence for the tumor class, forming a pool of tumor-positive tiles, and the same for non-tumor tissue classes, forming a pool of non-tumor tiles. The model is then retrained by drawing weakly annotated tiles from the two pools.
%This enables a more informed training as it reduces redundancy of uninformative image tiles. Moreover, since we currently assign the slide-level tumor detection label to each training tile, not all tissue in a tumor-positive slide is necessarily tumorous. Therefore, this could reduce erroneous labels and consequently improve the optimization space.

\newpage
%Bibliography
\bibliographystyle{unsrt}  
\bibliography{references}

\end{document}